\documentclass[aps,pra,10pt,twocolumn,superscriptaddress,showkeys,showpacs]{revtex4-1}

\usepackage{graphicx}
\usepackage{dcolumn}
\usepackage{bm}
\usepackage{amsmath}
\usepackage{amssymb}
\usepackage{amsbsy}
\usepackage{color,soul} 
\usepackage[normalem]{ulem} 
\usepackage{enumitem} 

\newcommand{\ket}[1]{\left|#1\right\rangle}

\newcommand{\erfc}{\textrm{erfc}}
\newcommand{\erf}{\textrm{erf}}

\begin{document}


\title{Enhancing qubit readout through dissipative sub-Poissonian dynamics}

\author{B.~D'Anjou}
\affiliation{Department of Physics, McGill University, Montreal, Quebec, H3A 2T8, Canada}
\affiliation{Department of Physics, University of Konstanz, D-78464 Konstanz, Germany}
\author{W.A.~Coish}
\affiliation{Department of Physics, McGill University, Montreal, Quebec, H3A 2T8, Canada}
\affiliation{Quantum Information Science Program, Canadian Institute for Advanced Research, Toronto, Ontario, M5G 1Z8, Canada}
\affiliation{Center for Quantum Devices, Niels Bohr Institute, University of Copenhagen, 2100 Copenhagen, Denmark}
\date{\today}

\begin{abstract}
Single-shot qubit readout typically combines high readout contrast with long-lived readout signals, leading to large signal-to-noise ratios and high readout fidelities. In recent years, it has been demonstrated that both readout contrast and readout signal lifetime, and thus the signal-to-noise ratio, can be enhanced by forcing the qubit state to transition through intermediate states. In this work, we demonstrate that the sub-Poissonian relaxation statistics introduced by intermediate states can reduce the single-shot readout error rate by orders of magnitude even when there is no increase in signal-to-noise ratio. These results hold for moderate values of the signal-to-noise ratio ($\mathcal{S} \lesssim 100$) and a small number of intermediate states ($N \lesssim 10$). The ideas presented here could have important implications for readout schemes relying on the detection of transient charge states, such as spin-to-charge conversion schemes for semiconductor spin qubits and parity-to-charge conversion schemes for topologically protected Majorana qubits.
\end{abstract}


\pacs{03.67.Lx, 42.50.Lc, 03.65.Ta}
\maketitle

\section{Introduction}

High-fidelity single-shot readout of qubits is highly desirable for quantum information processing, and is crucial for achieving quantum error correction~\cite{divincenzo2001} and measurement-only quantum computation~\cite{raussendorf2001,bonderson2008,karzig2017}. In the last decade, single-shot readout was realized experimentally for a large variety of promising qubit implementations. Prominent examples include superconducting qubits~\cite{lupascu2008,mallet2009,liu2014,hover2014,jeffrey2014,walter2017}, trapped-ion and trapped-atom qubits~\cite{hume2007,myerson2008,gehr2010,noek2013,harty2014}, nitrogen-vacancy center spin and charge qubits in diamond~\cite{jiang2009,neumann2010,robledo2011,dreau2013,shields2015,danjou2016}, and semiconductor spin qubits~\cite{elzerman2004,barthel2009,morello2010,pla2013,eng2015,harveycollard2017-2,nakajima2017}. In the near future, it should also be possible to perform readout of topologically protected Majorana qubits via parity-to-charge conversion~\cite{aasen2016,plugge2017}.

The fidelity of these readout schemes is typically limited by spurious transitions, such as qubit relaxation, that limit the lifetime of the readout signal (i.e., the characteristic duration for which the two qubit states can be distinguished). As a result, the integration time is bounded and the signal-to-noise ratio (SNR) remains finite. Therefore, researchers have dedicated considerable effort to increasing both the readout signal lifetime and the readout contrast~\cite{robledo2011,studenikin2012,jeffrey2014,mason2015,shields2015,nakajima2017,harveycollard2017-2,earnest2017} with the aim of improving SNR and boosting readout fidelity. 

One promising approach that can achieve either of these goals is to make the system transition through intermediate states. For instance, it was demonstrated that the effective readout signal lifetime for the optical readout of a nitrogen-vacancy center spin in diamond can be enhanced by forcing the system to undergo flip-flops between electron and nuclear spins before reaching the steady state~\cite{steiner2010}. Similarly, it was shown that both the readout signal lifetime and the readout contrast for singlet-triplet qubits in semiconductors can be significantly improved by mapping one of the qubit states to a long-lived metastable charge state that is easily detected by a nearby charge sensor~\cite{studenikin2012,mason2015}. These methods have since been used to bring the charge discrimination fidelities required for the readout of singlet-triplet qubits in silicon close to $99.9\%$, making them a serious candidate for fault-tolerant quantum computation~\cite{nakajima2017,harveycollard2017-2}.

In this article, we show that incoherent transitions through intermediate states can be used to significantly enhance readout fidelity even when they do not contribute to an increase in SNR. The key observation underpinning this result is that the times at which consecutive transition events occur are correlated. Consequently, the total time taken to transition through the intermediate states is not exponentially distributed as would be expected for a direct transition (see Fig.~\ref{fig:fig1}). Instead, the probability distribution of the total transition time becomes peaked as the number of intermediate states increases (see Fig.~\ref{fig:fig2}). This reflects the sub-Poissonian correlations resulting from a transition event being conditioned on the occurrence of the previous ones~\cite{korotkov2000}. The randomness of the relaxation process is thus reduced, leading to an improved fidelity even in the case where the SNR is left unchanged. In particular, we show that the error probability may be reduced by orders of magnitude with the addition of a small number of intermediate states. These results identify an important resource for engineering of qubit readout that has so far been under-appreciated.

The article is structured as follows. In Sec.~\ref{sec:qubitReadout}, we introduce the basic formalism necessary to analyze the fidelity of a single-shot qubit readout. In Sec.~\ref{sec:minimalModel}, we review a minimal physical model of qubit readout that captures the effect of relaxation in the absence of intermediate states. The core of the article is Sec.~\ref{sec:enhancement}, where we show that the addition of intermediate states in the relaxation process may enhance readout fidelity without requiring an improvement in SNR. We discuss the results in Sec.~\ref{sec:discussion} and we conclude in Sec.~\ref{sec:conclusion}.

\section{Qubit readout \label{sec:qubitReadout}}

\subsection{Readout apparatus and outcomes \label{sec:apparatusOutcomes}}

We consider a qubit with computational basis states $\ket{+}$ and $\ket{-}$. The qubit state is read out by making the qubit interact with some (typically macroscopic) readout apparatus. If the readout outcome $\mathcal{O}$ registered by the apparatus is strongly dependent on the qubit state, single-shot qubit readout is possible. The readout outcome $\mathcal{O}$ may take various forms. For example, readout of superconducting qubits is often achieved by discriminating between two possible quadratures of a microwave tone~\cite{lupascu2008,mallet2009,liu2014,hover2014,jeffrey2014,walter2017}, while readout of trapped-ion qubits and nitrogen-vacancy center spin or charge qubits in diamond is typically performed by differentiating between two fluorescence signals of distinct intensity~\cite{hume2007,myerson2008,gehr2010,noek2013,harty2014,jiang2009,neumann2010,robledo2011,dreau2013,shields2015,danjou2016}. Similarly, readout of semiconductor spin qubits is usually carried out by mapping the qubit states to charge states that produce different electrical currents in a nearby single-electron transistor, quantum dot, or quantum point contact~\cite{elzerman2004,barthel2009,morello2010,pla2013,eng2015,harveycollard2017-2,nakajima2017}. These charge sensors could also be used to read out topologically protected qubits via a parity-to-charge conversion mechanism~\cite{aasen2016,plugge2017}.

\subsection{Average readout error rate \label{sec:averageErrorRate}}

An ideal readout apparatus yields distinct readout outcomes when the qubit is prepared in either of the computational basis states. This means that the two probability distributions $P(\mathcal{O}|\pm)$ of the outcome conditioned on the initial state being $\ket{\pm}$ should not overlap. It can then be unambiguously decided which of the two computational basis states generated the readout outcome. Such a measurement is said to be accurate or sharp~\cite{wiseman2010}. In practice, however, the probability distributions $P(\mathcal{O}|+)$ and $P(\mathcal{O}|-)$ have a finite overlap. Consequently, there is a finite probability $\epsilon_{\pm}$ that the state $\ket{\pm}$ will be incorrectly identified as $\ket{\mp}$. We refer to the probabilities $\epsilon_{\pm}$ as the conditional single-shot readout error rates. The accuracy of the readout can then be quantified using the average single-shot readout error rate,
\begin{align}
	\epsilon = \frac{\epsilon_+ + \epsilon_-}{2}. \label{eq:averageErrorRate}
\end{align}
The average single-shot readout fidelity $F=1-\epsilon$ is also commonly quoted as a measure of readout accuracy. Note that in Eq.~\eqref{eq:averageErrorRate}, the conditional error rates $\epsilon_{\pm}$ are weighed equally. Thus, the quantity $\epsilon$ is the probability of error in the case where $\ket{+}$ and $\ket{-}$ occur with equal probability \emph{a priori}. Information processing protocols are often designed to yield such equal prior probabilities to extract a maximum of information from the measurement. In the general case, Eq.~\eqref{eq:averageErrorRate} still provides a useful figure of merit for the accuracy of readout that has the advantage of being independent of the particular application at hand.

\subsection{Maximum-likelihood decision \label{eq:maximumLikelihood}}

The average single-shot readout error rate $\epsilon$, Eq.~\eqref{eq:averageErrorRate}, is completely determined by the readout outcome distributions $P(\mathcal{O}|\pm)$. To relate $\epsilon$ to these distributions, it is necessary to specify a decision rule mapping the readout outcome $\mathcal{O}$ to one of the computational basis states. The decision rule that minimizes $\epsilon$, Eq.~\eqref{eq:averageErrorRate}, is known as a maximum-likelihood decision rule~\cite{kay1998}. A maximum-likelihood decision compares the likelihood ratio $\Lambda = P(\mathcal{O}|+)/P(\mathcal{O}|-)$ to unity: 
\begin{align}
	\Lambda > 1 \rightarrow \ket{+}, \;\;\; \Lambda < 1 \rightarrow \ket{-}. \label{eq:decisionRuleMLE}
\end{align}
The above decision rule minimizes Eq.~\eqref{eq:averageErrorRate} because it is a special case of a maximum \emph{a posteriori} decision rule~\cite{kay1998}. A maximum \emph{a posteriori} decision maximizes the posterior probability $P(\pm|\mathcal{O})$ over the qubit states. It is thus manifestly optimal. When the two qubit states are equally likely to occur, as was assumed in Eq.~\eqref{eq:averageErrorRate}, maximizing $P(\pm|\mathcal{O})$ leads to Eq.~\eqref{eq:decisionRuleMLE}.

It is often the case that the readout outcome $\mathcal{O}$ is a real number, such as a voltage or a current. Moreover, it is common that the distributions $P(\mathcal{O}|\pm)$ are well separated and intersect at a single value $\mathcal{O}=\mathcal{O}_{\textrm{th}}$, given by solving $P(\mathcal{O}_{\textrm{th}}|+)=P(\mathcal{O}_{\textrm{th}}|-)$. In that case, the decision rule of Eq.~\eqref{eq:decisionRuleMLE} reduces to:
\begin{align}
	\mathcal{O} > \mathcal{O}_{\textrm{th}} \rightarrow \ket{+}, \;\;\; \mathcal{O} < \mathcal{O}_{\textrm{th}} \rightarrow \ket{-}. \label{eq:decisionRuleThreshold}
\end{align}
Equation~\eqref{eq:decisionRuleThreshold} is the decision rule that will be used in the remainder of this article.

\section{Minimal readout model \label{sec:minimalModel}}

In Sec.~\ref{sec:qubitReadout}, we gave a general description of a single-shot qubit readout. We now review a minimal model of qubit readout~\cite{gambetta2007} that incorporates the main limiting sources of error in a wide variety of implementations, namely, readout noise and qubit relaxation. Despite its simplicity, this model and its extensions~\cite{myerson2008,danjou2014,ng2014,danjou2016,danjou2017} have been used as the starting point for the analysis of qubit readout in various experiments~\cite{myerson2008,barthel2009,danjou2016,harveycollard2017-2,walter2017}. The formalism of the present section will be used as a benchmark to quantify the improvements discussed in Sec.~\ref{sec:enhancement}.

\begin{figure}
\includegraphics[width=\columnwidth]{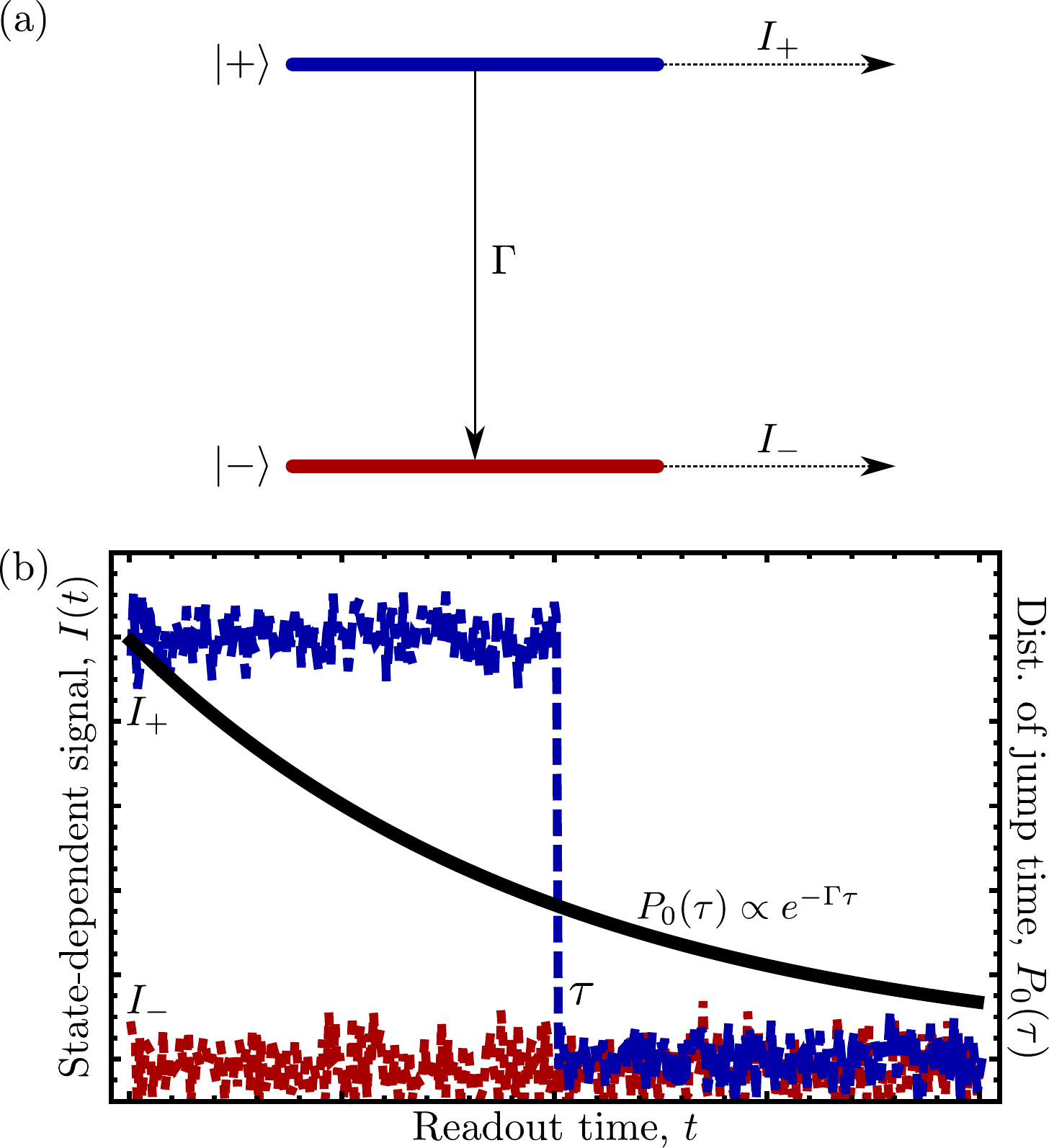}
\caption{Minimal model of qubit readout. (a) Level diagram of the minimal model. The detector registers a signal $I(t) = I_{\pm} + \delta I(t)$ when the qubit state is $\ket{\pm}$ at time $t$. Here, $\delta I(t)$ is Gaussian white noise. The excited state $\ket{+}$ relaxes to the ground state with probability per unit time $\Gamma$. (b) Instances of the readout signal $I(t)$ for both qubit states. The average readout signal for $\ket{-}$ (dotted red line) is a constant $I_-$ for all readout times $t>0$. The average readout signal for $\ket{+}$ (dashed blue line) is initially $I_+$ but jumps to $I_-$ at a random time $\tau>0$ due to relaxation. The probability distribution $P_0(\tau)$ of the jump time $\tau$ is exponential (solid black line). \label{fig:fig1}}
\end{figure}

The readout dynamics of the minimal model are illustrated in Fig.~\ref{fig:fig1}(a). If the qubit is in the ground state $\ket{-}$ at time $t=0$, it remains in the ground state for all times $t>0$. However, if the qubit is in the excited state $\ket{+}$, it relaxes to $\ket{-}$ at some random time $\tau>0$. We assume that the relaxation is an incoherent Markov process. The time $\tau$ is then distributed according to the exponential distribution $P_0(\tau) = \Gamma e^{-\Gamma \tau}$, where $\Gamma$ is the inverse relaxation time. The distribution $P_0(\tau)$ is illustrated in Fig.~\ref{fig:fig1}(b).

The qubit is coupled to a detector (e.g., a proximal charge sensor for readout of semiconductor spin qubits) that can distinguish between the two qubit states at any given time. More precisely, the detector registers a signal $I(t) = I_{\pm} + \delta I(t)$ if the qubit state is $\ket{\pm}$ at time $t$. Here, $I_{\pm}$ are state-dependent average detector signals and $\delta I(t)$ is detector noise. Without loss of generality, we assume $I_+ > I_-$ and we define the readout contrast $\Delta I = I_+ - I_-$. When the qubit transitions between different states, the transition appears as a sharp `jump' in the detector signal at time $\tau$~\footnote{For simplicity, we assume that the detector response time is much shorter than the relaxation time $\Gamma^{-1}$}. Particular instances of the readout signals $I(t)$ for both qubit states are illustrated in Fig.~\ref{fig:fig1}(b). After a transition occurs, the two qubit states can no longer be distinguished by the detector. Hence, the lifetime of the readout signal is given by the relaxation time $\left<\tau\right> = \Gamma^{-1}$.

If the detector signal were not noisy, the qubit state could be identified with perfect accuracy by integrating the signal for short times $t \ll \Gamma^{-1}$, thus avoiding relaxation events that would cause readout errors. In practice, however, noise in the detector sets a minimum readout time that limits the fidelity. For simplicity, we assume Gaussian white noise $\delta I(t)$ with autocorrelation function $\left<\delta I(t) \delta I(t')\right> = R^{-1} \delta(t-t')$. In this case, the two average signals $I_\pm$ can be resolved in a readout time $t$ of order $1/r$, where $r = R (\Delta I/2)^2$ is the rate of change of the (power) SNR. When $r \gg \Gamma$, the probability of a relaxation event occurring during that time is thus approximately $\Gamma/r$. Therefore, we expect the average error rate $\epsilon$ in that limit to be of order $1/\mathcal{S}$, where
\begin{align}
  \mathcal{S} = \frac{r}{\Gamma} = \frac{R \Delta I^2}{4\Gamma} \label{eq:SNR}
\end{align}
is the SNR obtained upon averaging the noise over a time $\Gamma^{-1}$. In the following, ``the SNR'' always refers to the quantity $\mathcal{S}$.

To make the above intuition more formal, we need to specify the measurement outcomes $\mathcal{O}$ introduced in Sec.~\ref{sec:qubitReadout}. A natural (although suboptimal~\cite{gambetta2007}) choice that captures the tradeoff between noise averaging and relaxation is the time-averaged signal $\bar{I}$:
\begin{align}
	\mathcal{O} = \bar{I} = \frac{1}{t} \int_{0}^{t} dt'\, I(t'). \label{eq:averageSignal}
\end{align}
Here, the readout time $t$ must be chosen to optimize the aforementioned tradeoff. Analytical expressions for the probability distributions $P(\bar{I}|\pm)$ as well as the corresponding single-shot readout error rate $\epsilon$ have been obtained~\cite{gambetta2007}. In particular, it can be shown that
\begin{align}
	\epsilon \sim \frac{1}{2 \mathcal{S}} \ln \mathcal{S}, \label{eq:minimalExpression}
\end{align}
as $\mathcal{S} \rightarrow \infty$. Here, ``$\sim$'' denotes the asymptotic equality. This result is in qualitative agreement with the heuristic argument given above, up to a numerical prefactor and logarithmic corrections.

\section{Readout enhancement via sub-Poissonian dynamics \label{sec:enhancement}}

In the previous section, the single-shot readout error rate was limited by random relaxation events from $\ket{+}$ to $\ket{-}$ at short readout times $t$. It follows that a natural strategy to improve readout is to suppress the frequency of these relaxation events. One way to achieve this would be to increase the relaxation time $\left<\tau\right> = \Gamma^{-1}$. In this section, however, we show that these errors can instead be suppressed by using intermediate states to engineer sub-Poissonian dynamics that reduce the variance of $\tau$, thus making the relaxation process more deterministic.

\begin{figure}
\includegraphics[width=\columnwidth]{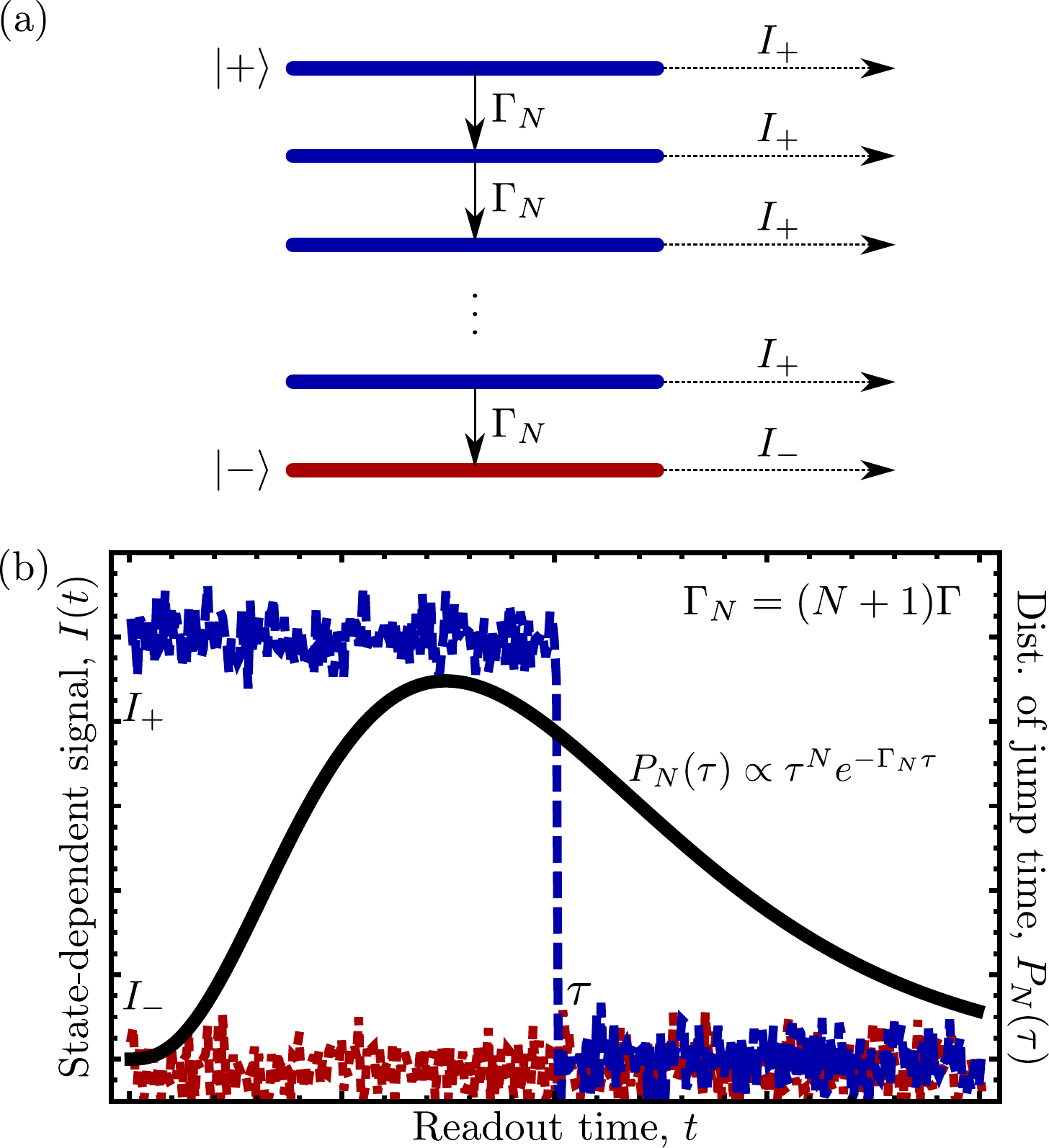}
\caption{Model of qubit relaxation through $N$ intermediate states. (a) The system transitions from $\ket{+}$ to $\ket{-}$ by cascading through the intermediate states. All transitions in the sequence occur with probability per unit time $\Gamma_N = (N+1)\Gamma$. It is assumed that all states except $\ket{-}$ yield the same average readout signal $I_+$. (b) Instances of the readout signal for both qubit states. As before, the average readout signal for $\ket{-}$ (dotted red line) is a constant $I_-$. The average readout signal for $\ket{+}$ (dashed blue line) is initially $I_+$ but jumps to $I_-$ at a time $\tau > 0$ when the system transitions from the last intermediate state to $\ket{-}$. Contrary to Fig.~\ref{fig:fig1}, however, the jump time $\tau$ now follows the Gamma distribution $P_N(\tau)$ (solid black line). \label{fig:fig2}}
\end{figure}

We introduce $N$ intermediate states between the excited state $\ket{+}$ and the ground state $\ket{-}$. The dynamics are such that the system must sequentially transition through all intermediate states in order to go from state $\ket{+}$ to state $\ket{-}$. This process is depicted in Fig.~\ref{fig:fig2}(a). Such cascaded dynamics can be approximately realized in existing qubit implementations~\cite{steiner2010,studenikin2012,mason2015,harveycollard2017-2,nakajima2017,gasparinetti2017} and were investigated in the context of transport through a chain of quantum dots~\cite{korotkov2000}.

We assume that all intermediate states couple to the detector as strongly as the excited state $\ket{+}$, i.e., the detector registers an average signal $I_+$ for all states except $\ket{-}$. Thus, a jump in the readout signal occurs at the time $\tau$ when the system transitions from the last intermediate state to $\ket{-}$ [see Fig.~\ref{fig:fig2}(b)]. Note also that the readout contrast remains unchanged from Sec.~\ref{sec:minimalModel}, $\Delta I = I_+ - I_-$. In addition, we assume that every transition along the sequence occurs with the same probability per unit time $\Gamma_N = (N+1) \Gamma$. As a result of the factor $N+1$, the lifetime of the readout signal is still $\left<\tau\right> = \Gamma^{-1}$.

Because both the readout contrast and the readout signal lifetime are the same as in Sec.~\ref{sec:minimalModel}, the SNR is again given by Eq.~\eqref{eq:SNR}. It might therefore be naively expected that the readout fidelity is not significantly affected. However, the introduction of intermediate states qualitatively modifies the statistics of the jump time $\tau$, which now follows the Gamma distribution
\begin{align}
  P_N(\tau) = \frac{\Gamma_{N}^{N+1} \tau^{N}}{N!} e^{-\Gamma_{N} \tau}. \label{eq:gammaDistribution}
\end{align}
The distribution $P_N(\tau)$ is illustrated in Fig.~\ref{fig:fig2}(b) for $N=3$. The distribution becomes peaked around $\tau = \Gamma^{-1}$, which strongly suppresses the probability that a jump event occurs at times smaller than $\Gamma^{-1}$. Indeed, according to Eq.~\eqref{eq:gammaDistribution}, the probability of a jump occurring within a readout time $t$ of order $r^{-1}$ is now roughly $(\Gamma_N/r)^{N+1}/(N+1)!$ provided that $r/\Gamma_N = \mathcal{S}/(N+1) \gg 1$. In this regime, we therefore expect the single-shot readout error rate to be proportional to:
\begin{align}
  \epsilon \propto \frac{1}{(N+1)!}\left(\frac{N+1}{\mathcal{S}}\right)^{N+1}. \label{eq:enhancedExpressionHeuristic}
\end{align}
For large enough $\mathcal{S}$, increasing the number of intermediate states $N$ should therefore lead to an exponential suppression of the error rate compared to the result of Sec.~\ref{sec:minimalModel} for $N=0$. A direct asymptotic expansion gives the correct leading-order asymptotic contribution to the error rate:
\begin{align}
  \epsilon \sim \frac{1}{2(N+1)!}\left\{\frac{N+1}{\mathcal{S}} \ln \left[2^N N! \left(\frac{\mathcal{S}}{N+1}\right)^{N+1}\right] \right\}^{N+1} \label{eq:enhancedExpression}
\end{align}
as $\mathcal{S} \rightarrow \infty$. Equation~\eqref{eq:enhancedExpression} is in qualitative agreement with the heuristic order-of-magnitude estimate, Eq.~\eqref{eq:enhancedExpressionHeuristic}, again up to a numerical prefactor and logarithmic corrections.

\begin{figure}
\includegraphics[width=\columnwidth]{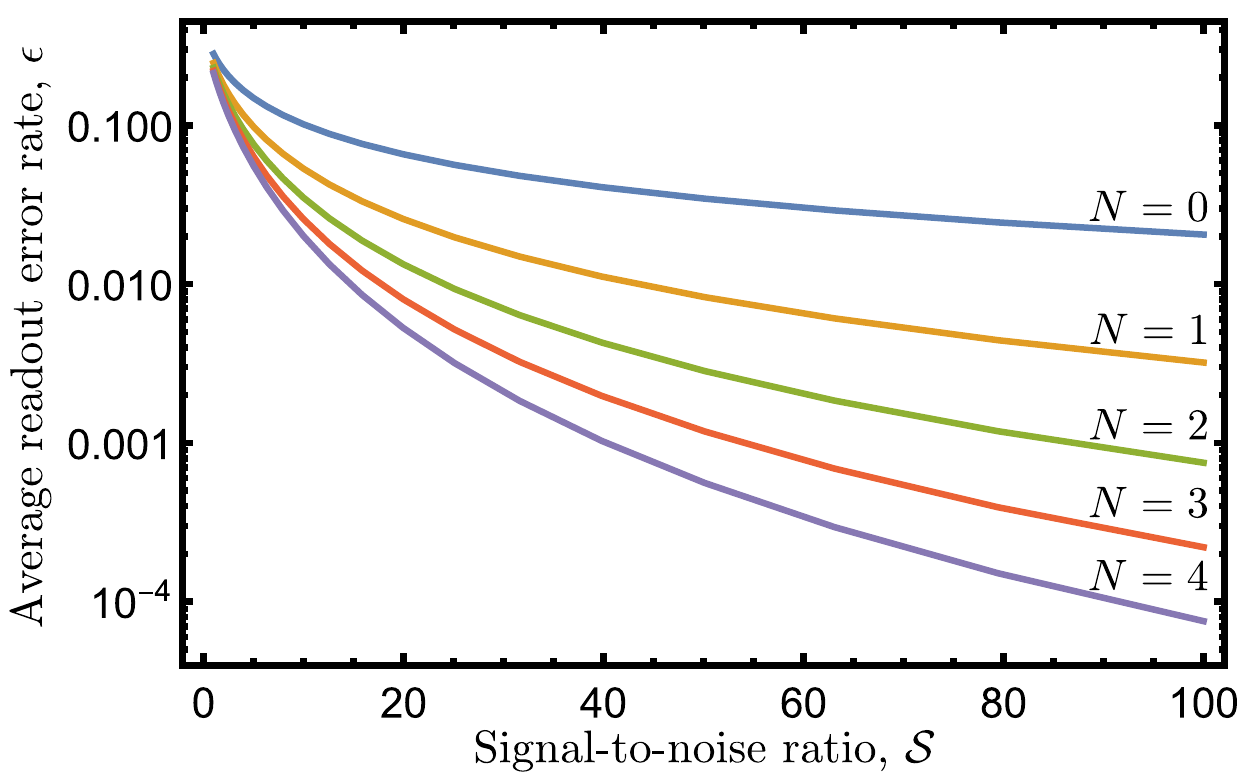}
\caption{Average single-shot readout error rate, Eq.~\eqref{eq:averageErrorRate}, on a logarithmic scale as a function of the SNR for different numbers $N$ of intermediate states. The reported values of $\epsilon$ are calculated as described in the main text and as detailed in Appendix~\ref{app:analytical}. \label{fig:fig3}}
\end{figure}

To verify that an advantage exists in a non-asymptotic parameter regime, we perform a full analysis of the statistics of the time-averaged signal $\bar{I}$, Eq.~\eqref{eq:averageSignal}, in Appendix~\ref{app:analytical}. Analytical expressions for the single-shot readout error rate $\epsilon$ are first derived in the presence of intermediate states. The error rate $\epsilon$ is then minimized numerically with respect to the readout time $t$ and decision threshold $\bar{I}_{\textrm{th}}$. We plot $\epsilon$ on a logarithmic scale as a function of $\textrm{SNR}$ for $N$ ranging from $0$ to $4$ in Fig.~\ref{fig:fig3}. For moderate values of the SNR ($\mathcal{S} \lesssim 100$), the addition of a single intermediate state leads to a near order-of-magnitude reduction in $\epsilon$.

Our analysis of the time-averaged signal $\bar{I}$, Eq.~\eqref{eq:averageSignal}, has enabled us to analytically and efficiently calculate the single-shot readout error rate, $\epsilon$. However, as mentioned in Sec.~\ref{sec:minimalModel}, the time-averaged signal is a suboptimal statistic for qubit readout. This is because time-averaging the signal discards single-shot information about the statistics of the jump time~\cite{gambetta2007}. In contrast, optimal readout is achieved by choosing the readout outcome to be the full state-dependent signal, $\mathcal{O} = I(t)$. Because time averaging is suboptimal, it is prudent to verify that the advantage discussed above is intrinsic to the system dynamics and not merely an artifact of the processing method. We therefore implemented the optimal processing method~\cite{ng2014,danjou2017} for the readout dynamics of Fig.~\ref{fig:fig2}(a). The relevant formalism is outlined in Appendix~\ref{app:optimal}. We use this method to obtain a Monte Carlo estimate of the minimum theoretically achievable error rate. The results are illustrated in Fig.~\ref{fig:fig5} of Appendix~\ref{app:optimal} for $\mathcal{S} = 20$ and several values of $N$. While the optimal method reduces the error rate compared to the time-averaging method, the readout enhancement due to sub-Poissonian dynamics remains the same.

Our results were obtained under the assumption that the detector registers the same contrast $\Delta I$ for all states except $\ket{-}$. Moreover, it was assumed that all transitions occur at the same rate $\Gamma_N = (N+1)\Gamma$. In practice, however, there will necessarily be variations in the contrast and in the transition rate at each step of the cascade. Therefore, it is important to verify that the readout enhancement persists in spite of small asymmetries in these parameters. Using the optimal method described in Appendix~\ref{app:optimal}, we show numerically that the enhanced error rate is weakly dependent on either an asymmetry in contrast or an asymmetry in transition rates. The results are shown in Fig.~\ref{fig:fig6} of Appendix~\ref{app:asymmetry} for $N=1$ and $\mathcal{S} = 20$. Thus, no fine-tuning is required for the readout to benefit from the sub-Poissonian dynamics.

\section{Discussion \label{sec:discussion}}

The results presented here show that intermediate states can be useful even in situations where the SNR is unchanged. This means that the SNR is in general not sufficient to fully characterize the readout. We thus identify sub-Poissonian qubit dynamics as an \emph{additional resource} for qubit readout engineering. Depending on the experimental constraints, either or both of these resources can be used to optimize readout fidelity.

In Sec.~\ref{sec:enhancement}, for instance, we fix the readout signal lifetime $\left<\tau\right> = \Gamma^{-1}$ by increasing the transition rates proportionally to $N$, $\Gamma_N = (N+1)\Gamma$. This is done to demonstrate that sub-Poissonian statistics of the jump time $\tau$ may reduce the error rate without increasing SNR. In general, however, the addition of intermediate states may increase $\left<\tau\right>$ (and thus the SNR) as well as reduce the randomness in $\tau$. As can be seen in Fig.~\ref{fig:fig3}, such a combined increase in SNR and $N$ would lead to an even greater suppression of errors. In particular, the enhancement of the readout of the $NV$-center spin in diamond reported in Ref.~\cite{steiner2010} should already benefit from this combination of effects even though high-fidelity single-shot readout has not been achieved.

\begin{figure}
\includegraphics[width=\columnwidth]{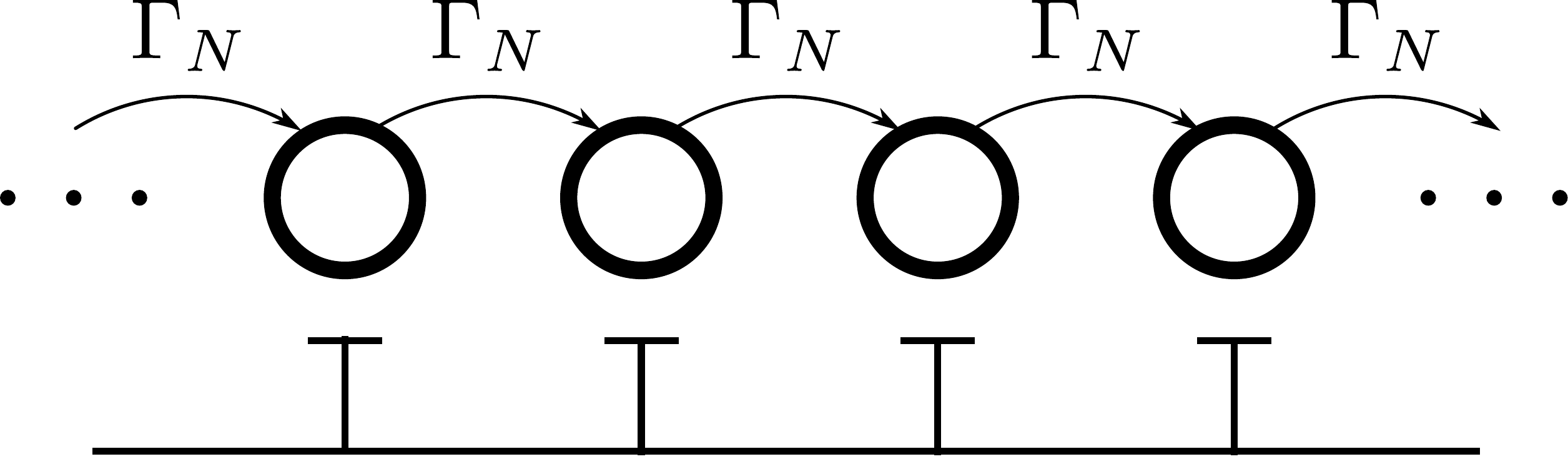}
\caption{Sub-Poissonian charge dynamics in a chain of quantum dots~\cite{korotkov2000}. Metallic gates are used to maintain strong capacitive coupling between distant dots while the charge hops from one quantum dot to the next (circles) with probability per unit time $\Gamma_N$. The resulting statistics for the total transition time $\tau$ are given by Eq.~\eqref{eq:gammaDistribution}. \label{fig:fig4}}
\end{figure}

Readout of semicondutor spin qubits~\cite{elzerman2004,barthel2009,morello2010,pla2013,eng2015,harveycollard2017-2,nakajima2017} and recent proposals for readout of topologically protected qubits~\cite{aasen2016,plugge2017} frequently rely on the detection of a transient charge state (or a transient flow of charge) in quantum dots. In these schemes, charge relaxation often limits the readout signal lifetime. Stabilizing the charge state is therefore crucial to improving the readout fidelity. Engineering sub-Poissonian charge dynamics could be a promising avenue for achieving this goal. For instance, the relaxation process could be made generically sub-Poissonian by forcing the charge to tunnel through a chain of quantum dots in the Coulomb blockade regime~\cite{korotkov2000}. In principle, such chains can be made long enough to include several dots by using gates to enforce long-range capacitive coupling between the dots~\cite{chan2002,hu2007,flensberg2010,trifunovic2012}. Such a long-range coupling prevents double occupancy of the chain, leading to highly sub-Poissonian statistics of the total transition time (see Fig.~\ref{fig:fig4}).

Although we restricted our attention to a cascade of incoherent transitions, the engineering of sub-Poissonian relaxation statistics could, in principle, improve readout for more general dynamics including, e.g., coherent dynamics described by a Lindblad quantum master equation. The only requirement is that the distribution $P(\tau)$ becomes narrow enough to suppress the probability of a jump occurring during the readout. More generally, it should only be necessary to make $\tau$ more deterministic by reducing the (differential) entropy of $P(\tau)$.

\section{Conclusion \label{sec:conclusion}}

In summary, we show that sub-Poissonian dynamics is a useful resource for qubit readout engineering. In particular, we demonstrate that forcing a qubit to relax through a sequence of intermediate states can reduce the single-shot readout error rate by orders of magnitude. This enhancement is entirely due to the induced sub-Poissonian statistics of the relaxation process and thus holds even when the SNR is unchanged. We have also verified that the enhancement is significant for moderate values of the SNR ($\mathcal{S} \lesssim 100$) and small numbers of intermediate states ($N \lesssim 10$). The example given here makes it clear that the single-shot readout fidelity should be optimized accounting for the full readout statistics, rather than simply quoting the SNR.

A particularly promising application of our results is the enhanced detection of short-lived charge states using charge sensors. As discussed in Sec.~\ref{sec:discussion}, charge relaxation dynamics can be made generically sub-Poissonian with the help of quantum dot chains. This could have important implications for recent proposals to read out topologically protected Majorana qubits via parity-to-charge conversion.

\section*{Acknowledgments}

We acknowledge financial support from the Natural Sciences and Engineering Research Council of Canada (NSERC), the Canadian Institute for Advanced Research (CIFAR), the Institut Transdisciplinaire d'Information Quantique (INTRIQ), and Nordea Fonden.

\appendix

\section{Analytical expressions for the single-shot readout error rate \label{app:analytical}}

In this Appendix, we briefly outline the derivation of the analytical expressions used to plot Fig.~\ref{fig:fig3}. The formalism presented here may also be used as the starting point for the derivation of asymptotic results such as Eqs.~\eqref{eq:minimalExpression} and \eqref{eq:enhancedExpression}.

The probability distribution of the time-averaged signal $\bar{I}$, Eq.~\eqref{eq:averageSignal}, depends on the qubit state. When the qubit state is $\ket{-}$, the average detector signal is $I_-$ for all readout times $t>0$. Assuming Gaussian white noise of the form $\left<\delta I(t) \delta I(t')\right> = R^{-1} \delta(t-t')$, $\bar{I}$ is a Gaussian random variable with mean $I_-$ and variance $1/Rt$:
\begin{align}
  P(\bar{I}|-) = \mathcal{N}\left(\bar{I};I_-,1/\sqrt{R t}\right),
\end{align}
where
\begin{align}
  \mathcal{N}(x;\mu,\sigma) = \frac{1}{\sqrt{2\pi \sigma^2}} e^{-\frac{(x-\mu)^2}{2\sigma^2}}
\end{align}
is the normal distribution of mean $\mu$ and variance $\sigma^2$. When the qubit state is $\ket{+}$, the presence of relaxation leads to a non-Gaussian distribution $P(\bar{I}|+)$. However, the distribution $P(\bar{I}|+,\tau)$ conditioned on a particular value of the jump time $\tau$ is Gaussian. Using Bayes' rule, we have:
\begin{align}
  P(\bar{I}|+) = \int_{0}^{\infty} d\tau P(\bar{I}|+,\tau) P_N(\tau), \label{eq:excitedStateBayes}
\end{align}
where
\begin{align}
  P(\bar{I}|+,\tau) =
	\left\{
	\begin{array}{ll}
	\mathcal{N}\left(\bar{I};I_- + (\tau/t) \Delta I,1/\sqrt{R t}\right) & \textrm{if}\;\tau < t \\
	\mathcal{N}\left(\bar{I}; I_+,1/\sqrt{R t}\right) & \textrm{if}\;\tau \ge t
	\end{array}
	\right.
\end{align}
According to the decision rule outlined in Sec.~\ref{sec:qubitReadout}, the conditional error rates $\epsilon_\pm^{(N)}$ for a given value of $N$ are given by the probabilities that $\bar{I}$ is above (below) the decision threshold $\bar{I}_\textrm{th}$ when the qubit state is $\ket{-}$ ($\ket{+}$):
\begin{align}
\begin{split}
  \epsilon_+^{(N)} &= \int_{-\infty}^{\bar{I}_{\textrm{th}}} d\bar{I}\,P(\bar{I}|+), \\
	 \epsilon_-^{(N)} &= \int_{\bar{I}_\textrm{th}}^{\infty} d\bar{I}\,P(\bar{I}|-). \label{eq:conditionalErrorRatesN}
\end{split}
\end{align}

The above integrals were performed in Ref.~\cite{gambetta2007} for the case $N=0$. The resulting analytical expressions for the conditional error rates $\epsilon_\pm^{(0)}$ are:
\begin{align} 
\begin{split}
  \epsilon_+^{(0)} &=\frac{1}{2}e^{\rho\frac{\gamma}{r}\left(\frac{\gamma}{8 r} - \nu\right)} \\
		 & \times \left\{ \erf\left[\sqrt{2\rho}\left(\frac{\gamma}{4 r} - \nu\right)\right] - \erf\left[\sqrt{2\rho}\left(\frac{\gamma}{4 r}+ 1 - \nu\right)\right] \right\} \\
		&+ \frac{1}{2}\erfc\left(-\sqrt{2 \rho \nu^2}\right), \label{eq:analyticN0} \\
		\epsilon_-^{(0)} &= \frac{1}{2}\erfc\left(\sqrt{2 \rho \nu^2}\right). 
\end{split}
\end{align}
Here, $r = R (\Delta I/2)^2$, $\rho = r t$, $\nu = (\bar{I}_{\textrm{th}} - I_-)/\Delta I$, and $\gamma$ is the probability per unit time to transition from one state to the next [set to $\gamma = \Gamma_N = (N+1)\Gamma$ in the main text].

To obtain the error rates for $N > 0$, we notice that
\begin{align}
  P_N(\tau) = \frac{(-1)^N \gamma^{N+1}}{N!} \frac{d^N}{d\gamma^N} \left(\frac{P_0(\tau)}{\gamma}\right),
\end{align}
where $P_0(\tau) = \gamma e^{-\gamma \tau}$ is the exponential distribution. It then follows from Eqs.~\eqref{eq:excitedStateBayes} and \eqref{eq:conditionalErrorRatesN} that analytical expressions for $\epsilon_\pm^{(N)}$ when $N>0$ can be obtained by taking derivatives of Eq.~\eqref{eq:analyticN0} with respect to the transition rate $\gamma$:
\begin{align}
\begin{split}
  \epsilon_+^{(N)} &= \frac{(-1)^N \gamma^{N+1}}{N!}\frac{d^N}{d\gamma^N} \left(\frac{\epsilon_+^{(0)}}{\gamma}\right), \\
	 \epsilon_-^{(N)} &= \epsilon_-^{(0)}.
\end{split}
\end{align}
These expressions are used to minimize the average single-shot readout error rate $\epsilon^{(N)} = \left[\epsilon_+^{(N)} + \epsilon_-^{(N)}\right]/2$ with respect to both $\rho$ and $\nu$. Setting $\gamma = (N+1) \Gamma$ and $r/\Gamma = \mathcal{S}$, this yields the results presented in Fig.~\ref{fig:fig3}.

\section{Comparison to optimal processing \label{app:optimal}}

The optimal processing method calculates the likelihood ratio $\Lambda$ from the full signal $\mathcal{O} = I(t)$. This is achieved by solving a set of linear classical It{\= o} stochastic differential equations for the unconditioned system state $\boldsymbol{\ell}$~\cite{ng2014,danjou2017}. These equations, also known as filtering equations, have the form:
\begin{align}
  d\boldsymbol{\ell} &= \left[\mathcal{L} + I(t) \boldsymbol{\mathcal{I}} R\right] \boldsymbol{\ell} dt \;\;\; \textrm{(It{\= o})}. \label{eq:ito}
\end{align}
Here, $\boldsymbol{\ell}$ is a vector with components $\ell_i$ such that the probability of finding the system in state $\ket{i}$ conditioned on past observations is $\ell_i/\sum_j \ell_j$. The matrix $\mathcal{L}$ is the rate matrix that describes the transitions illustrated in Fig.~\ref{fig:fig2}, while the matrix $\mathcal{I} = \textrm{diag}(I_0,I_1,\dots,I_{N+1})$ encodes the average signals $I_i$ registered by the detector for each state $\ket{i}$ along the cascade. The likelihood ratio is given by:
\begin{align}
  \Lambda = \frac{\sum_i \ell_i^{(+)}}{\sum_i \ell_i^{(-)}}, \label{eq:lambda}
\end{align}
where $\boldsymbol{\ell}^{(\pm)}$ is the solution of Eq.~\eqref{eq:ito} assuming an initial state $\ket{\pm}$. In practice, Eq.~\eqref{eq:ito} requires a very small time step to solve accurately. We therefore use an alternative set of ordinary differential equations~\cite{wilkie2004,gambetta2007,danjou2014}:
\begin{align}
  \frac{d\boldsymbol{\ell}}{dt} = \left[\mathcal{L} + \left(I(t) - \frac{1}{2}\boldsymbol{\mathcal{I}}\right) \boldsymbol{\mathcal{I}} R\right] \boldsymbol{\ell}. \label{eq:formal}
\end{align}
It can be shown that Eq.~\eqref{eq:formal} is equivalent to Eq.~\eqref{eq:ito} provided that it is solved with a higher-order numerical method such as the fourth-order Runge-Kutta method~\cite{press2002}. The use of a higher-order method enables an accurate computation of $\Lambda$ with a reasonable number of time steps. Note that Eq.~\eqref{eq:formal} may not be solved with the Euler method since its first-order increment differs from that of Eq.~\eqref{eq:ito}.

\begin{figure}
\includegraphics[width=\columnwidth]{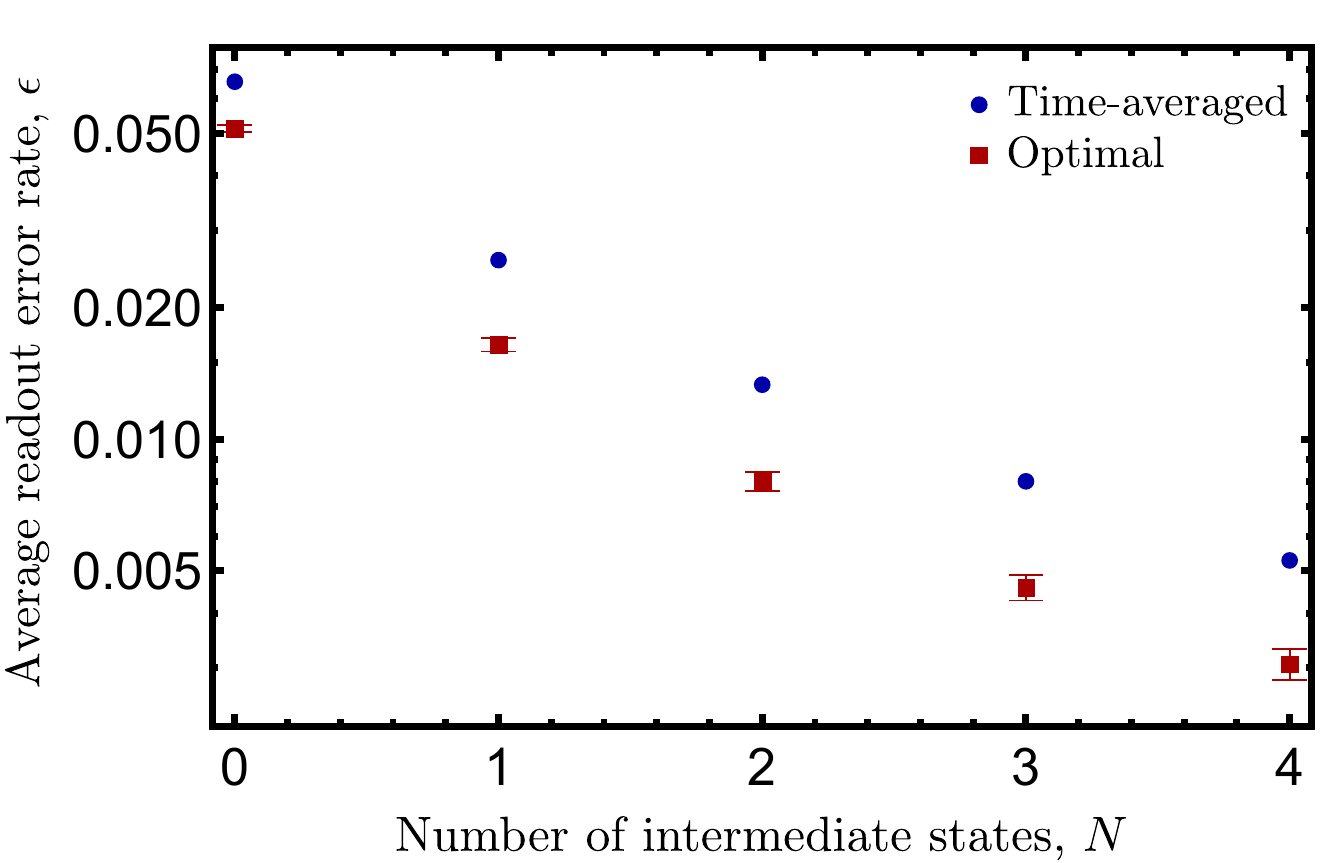}
\caption{Average error rate $\epsilon$ as a function of the number of intermediate states $N$ using the time-averaged method, Eq.~\eqref{eq:averageSignal} (blue dots), and using the optimal method, Eq.~\eqref{eq:formal} (red squares). The SNR is $\mathcal{S} = 20$. In the optimal case, each error rate is obtained by simulating $5\times 10^4$ signals $I(t)$ for each state $\ket{\pm}$. The error bars for the optimal method are given by the standard deviation of a binomial process with frequency $\epsilon$. \label{fig:fig5}}
\end{figure}

For both states $\ket{\pm}$, a large number of signals $I(t)$ are simulated. For a given set of parameters, the measurement time is chosen to be long enough to achieve the minimum error rate. For each signal, Eq.~\eqref{eq:formal} is solved using a fourth-order Runge-Kutta method. The likelihood ratio $\Lambda$ is then calculated using Eq.~\eqref{eq:lambda}. If $\Lambda < 1$ ($\Lambda > 1$) when the state is $\ket{+}$ ($\ket{-}$), an error is recorded. This procedure gives a Monte Carlo estimate of the conditional single-shot readout error rates $\epsilon_{\pm}$. These values are then used to calculate $\epsilon = (\epsilon_+ + \epsilon_-)/2$. The results are shown in Fig.~\ref{fig:fig5} for $\mathcal{S} = 20$ and several values of $N$.

\section{Asymmetrical parameters \label{app:asymmetry}}

\begin{figure}
\includegraphics[width=\columnwidth]{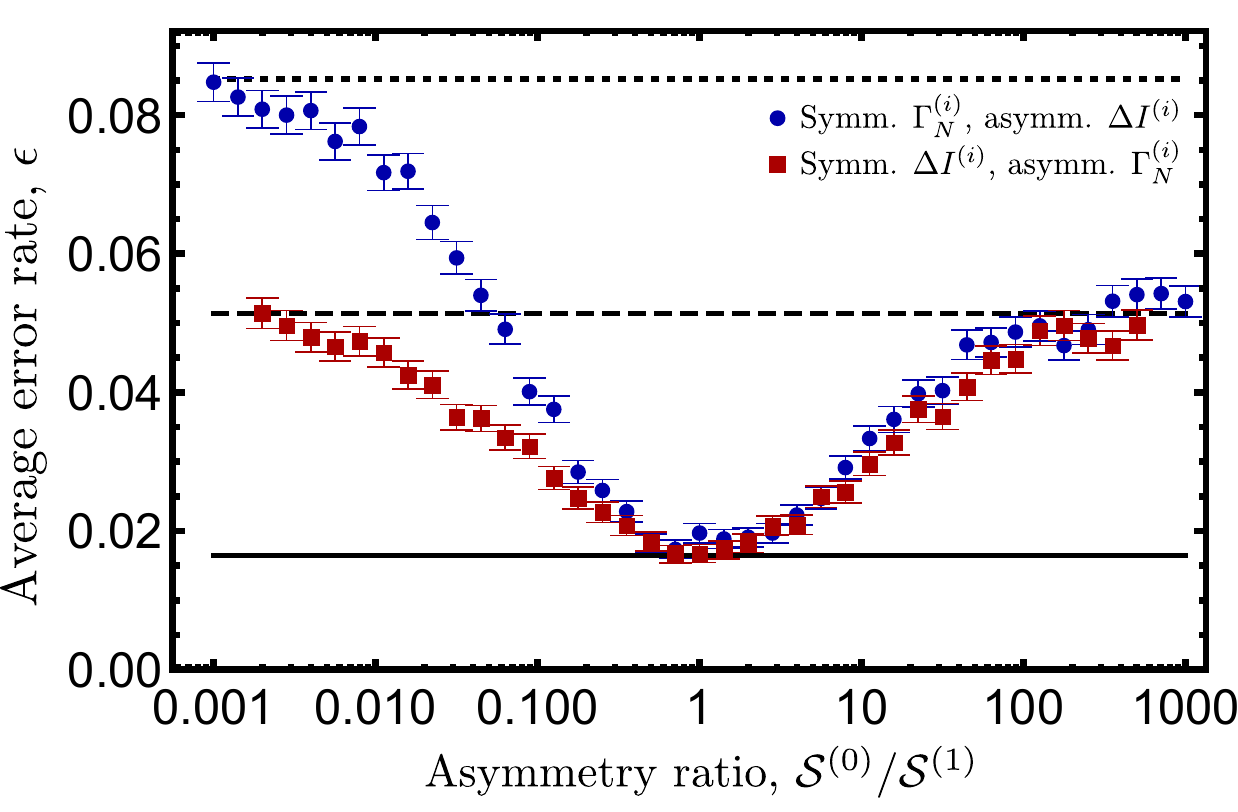}
\caption{Average error rate $\epsilon$ as a function of the asymmetry ratio $\mathcal{S}^{(0)}/\mathcal{S}^{(1)}$ in the case of $N=1$ intermediate state. The asymmetry in SNR arises from an asymmetry in contrast $\Delta I^{(i)}$ (blue dots) or from an asymmetry in transition rates $\Gamma_N^{(i)}$ (red squares). The total SNR is $\mathcal{S}=\mathcal{S}^{(0)}+\mathcal{S}^{(1)}=20$. Each error rate is obtained by simulating $10^4$ signals $I(t)$ for each state $\ket{\pm}$ and then applying the optimal processing method, Eqs.~\eqref{eq:lambda} and \eqref{eq:formal}. The error bars are given by the standard deviation of a binomial process with frequency $\epsilon$. The horizontal lines are the error rates obtained from simulations of the symmetric case (solid) and completely asymmetric cases (dashed and dotted) with $10^5$ samples for each state. The completely asymmetric cases correspond to the readout dynamics discussed in Refs.~\cite{gambetta2007,danjou2014}. \label{fig:fig6}}
\end{figure}

In this Appendix, we discuss the effect of an asymmetry in contrast or in transition rates. A convenient way to parametrize the asymmetry is to define partial SNRs, $\mathcal{S}^{(i)} = R \Delta {I^{(i)}}^2/4 \Gamma_N^{(i)}$, for $i=0,1,\dots,N$. Here, $\Delta I^{(i)}$ and $\Gamma_N^{(i)}$ are the $i^{\textrm{th}}$ contrast and transition rate in the cascade, respectively. The total $\textrm{SNR}$ is then defined as $\mathcal{S} = \sum_i \mathcal{S}^{(i)}$. For a fair comparison of the asymmetric case with the symmetric case, $\mathcal{S}$ must remain fixed as the $\mathcal{S}^{(i)}$ are varied. When the $\mathcal{S}^{(i)}$ differ significantly from each other, the advantage due to sub-Poissonian dynamics is expected to disappear since only a minority of intermediate states are responsible for most of the $\textrm{SNR}$. To demonstrate the resilience of the sub-Poissonian enhancement to an asymmetry in parameters, we use the optimal method of Appendix~\ref{app:optimal} to calculate the error rate as a function of the asymmetry ratio $\mathcal{S}^{(0)}/\mathcal{S}^{(1)}$ for the case $N=1$ and $\mathcal{S} = 20$. We introduce an asymmetry
\begin{enumerate}[label= \arabic*)]
\item by choosing symmetric $\Gamma_N^{(i)}$ and varying the $\Delta I^{(i)}$, and
\item by choosing symmetric $\Delta I^{(i)}$ and varying the $\Gamma_N^{(i)}$.
\end{enumerate}
The results are shown in Fig.~\ref{fig:fig6}. In either case, the symmetry point is a minimum. Moreover, the error rate does not change significantly provided that the asymmetry ratio $\mathcal{S}^{(0)}/\mathcal{S}^{(1)}$ differs from unity by less than an order of magnitude. For large asymmetries, we recover the readout dynamics discussed in Refs.~\cite{gambetta2007,danjou2014}, both of which lead to higher error rates than in the symmetric case.

\clearpage

\bibliography{pra.02}


\end{document}